\documentstyle[twoside,fleqn,espcrc2,epsf]{article}

\newcommand{\bee}{\begin{equation}}
\newcommand{\ee}{\end{equation}}
\newcommand{\beea}{\begin{eqnarray}}
\newcommand{\eea}{\end{eqnarray}}

\newcommand{\AmS}{{\protect\the\textfont2
  A\kern-.1667em\lower.5ex\hbox{M}\kern-.125emS}}

\title{Fixed point action and topological charge for SU(2) gauge
theory }

\author{ Thomas A. DeGrand\address{Physics Department,
                   University of Colorado,
                   Boulder, CO 80309}, %
          Anna Hasenfratz$^{\rm a}$,
\thanks{ Talk presented by Anna Hasenfratz}
         Decai Zhu$^{\rm a}$ }

\begin{document}

\begin{abstract}
We present a theoretically consistent definition of the topological
charge operator based on renormalization group arguments. 
Results of the measurement 
of the topological susceptibility at zero and finite temperature for
SU(2) gauge theory are presented.
\end{abstract}

\maketitle

\section{INTRODUCTION}

Instantons in the QCD vacuum can explain the $U_A(1)$ problem, and 
they
could be responsible for the low energy hadron and glueball spectrum
and the chiral phase transition  as well\cite{SHUR_lat96}. Instantons have been
studied on the lattice for over a decade but these studies have produced
about as much controversy as physical results. The problems are
three-fold: 1) Instantons are not conserved by topology on the lattice
- small instantons can "fall through" the lattice. 2) Most lattice
actions are not scale invariant -
the lattice action of an instanton depends on its size. 3) The
definition of the topological charge is problematic - the geometric
definition  erroneously identifies dislocations as instantons, while
the algebraic definition relies on some sort of smoothing  algorithm
like cooling or APE-smearing and the topological susceptibility has
both additive and multiplicative renormalization. 
In figure 1 the action of a single smooth instanton calculated with
the  Wilson action  is plotted as a
function of its size $\rho$. ($S_I=8\pi^2$ is the continuum
instanton action.) The dotted vertical line indicates where the
instanton disappears from the lattice according to the geometric
definition and the dashed line is the value of the topological charge
evaluated with the simplest algebraic $F\tilde F$ operator, the twisted
plaquette. The scale violation of the Wilson action is well known but 
it is interesting to note that the algebraic definition
gives $Q< 0.8$ if $\rho<2.0$ even for very smooth instantons.

\begin{figure}[htb]
\begin{center}
\vskip -20mm
\leavevmode
\epsfxsize=60mm
\epsfbox[40 50 530 590]{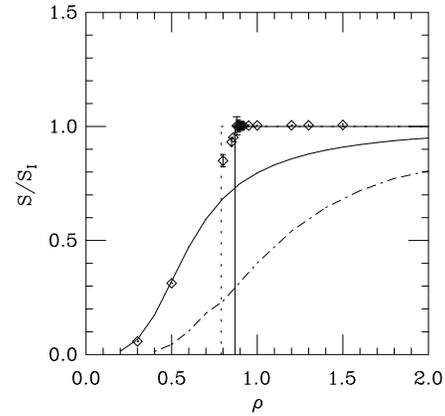}
\vskip -10mm
\end{center}
\caption{ Action of single 
instanton configurations computed using the Wilson action (solid curve).
The diamond symbols show the FP action value calculated according to
Eqn. 3.
The dotted vertical line indicates the
$Q=0 \to Q=1$ boundary calculated on the original coarse lattice 
with the geometric definition while the solid vertical 
line is the
boundary on the inverse blocked configuration. The dashed line is
$Q_{F\tilde F}$. }
\label{fig:sbc1w}
\end{figure}

Fixed point actions are, by construction, scale invariant, solving
problem 2). In addition, 
the renormalization group equation of fixed point actions
offer a theoretically consistent definition of the topological charge.
Since fixed point actions are 1-loop perfect but do not follow exactly
the
renormalized trajectory, this method does not solve 1). 

In this paper we describe the  renormalization group definition of the
topological charge $Q_{RG}$ and discuss our numerical results for the
topological susceptibility both at zero and at finite temperature. Further
details and an extensive reference list can be found in 
Ref. \cite{INSTANTON}.

\section{FIXED POINT ACTIONS AND THE DEFINITION OF $Q_{RG}$ }

Fixed point (FP) actions  of SU(N) gauge theories 
are 1-loop perfect actions corresponding to
the  $\beta=\infty$ fixed point of a renormalization group
transformation.  The defining renormalization group equation is 
\bee
S^{FP}(V)=\min_{ \{U\} } \left( S^{FP}(U) +T(U,V)\right),
\label{STEEP}
\ee
where $U$ is the original link variable, $V$ is the blocked link
variable and $T(U,V)$ is the blocking kernel that defines the
transformation. Eqn.~\ref{STEEP} defines, in a unique way,
 a fine lattice $\{U\}$ for every 
 coarse configuration  $\{V\}$ but it is not a simple interpolation of the
coarse lattice.  
The fine lattice $\{U\}$ blocks into  $\{V\}$
under the renormalization group transformation but it is  special 
among the  
many configurations that block into the original coarse
lattice
as it is the smoothest one. 
Indeed, for typical coarse configurations with
correlation length of a few lattice spacings  and plaquette expectation
value about 0.5 out of one, the plaquette expectation
value on the
fine configuration is typically between 0.95 and 1.0.
 We will refer to the transformation of Eqn.~\ref{STEEP}
 as  ``inverse blocking''.

It can be proven that the inverse blocking does not change the
topological properties of the original lattice, but it changes the
scale of any topological objects by a factor of two, i.e. if the coarse
lattice had certain number of instantons and anti-instantons, they will
be present on the fine lattice but with size doubled. Because the fine
lattice is very smooth, the instantons are easily identified and one
can avoid 
most of the problems related to the definition of the topological
charge. We define the charge of a coarse configuration as the charge of
the inverse blocked configuration measured with any reliable method.
In the following we use the geometric definition
\bee
Q_{RG}(V) = Q_{geom}(U).
\ee
The minimization  equation gives the value of the fixed point action 
on the coarse lattice as well
\bee
S^{FP}(V) = S^{FP}(U(V))+T(U(V),V) \label{SINVB}.
\ee
As it is much easier to parametrize the FP action on the fine
configuration, we use this relation in testing the scale
invariance of the FP action.

In figure 1 the solid vertical line corresponds to the boundary
$Q_{RG}=0 \to Q_{RG}=1$ and the diamond symbols show the value of
the action calculated with the FP action according to Eqn.
\ref{SINVB}. Unlike the plaquette action, the FP action is consistent
with the theoretical continuum value, independent of the size of the
instanton as long as the instanton is present, $Q_{RG}=1$.That demonstrates
not only the scale invariance of the  FP action but the validity of the
definition of $Q_{RG}$. 

We would like to emphasize that the topological properties of any
configuration are a dynamical question; they are defined with respect to a
given action. Our definition works with a FP action and it cannot be
automatically applied to configurations generated with any other
action.

\section{ THE TOPOLOGICAL SUSCEPTIBILITY AT T=0} 

We have measured the topological susceptibility using $Q_{RG}$ with
an approximate
SU(2) FP action. 
 The action consists of several powers of
two loops, the plaquette and the perimeter-six loop (x,y,z,-x,-y,-z).
Its coefficients are tabulated in Ref. \cite{INSTANTON}.

This 8 parameter FP action shows almost perfect scaling properties for
couplings weaker than the $N_T=2$ finite temperature phase transition.
Since the inverse blocking transformation is quite computer intensive,
we have used relatively small lattices and
concentrated on the strong coupling region comparing identical
physical volume measurements to see scaling or the lack of it. 
We use the parameter  $r_0$,
as defined through the force, by $r_0^2F(r_0)= -1.65$ to set the scale
in our calculation. We calculate  the dimensionless ratio
of susceptibility  times the fourth power of $r_0$
\bee
{ \chi r_0^4} = {\langle Q^2 \rangle}({r_0 \over L})^4
\ee
on lattices with approximately identical physical volumes
characterized by the quantity $L/r_0$. 

\begin{figure}[htb]
\begin{center}
\vskip -20mm
\leavevmode
\epsfxsize=60mm
\epsfbox[40 50 530 590]{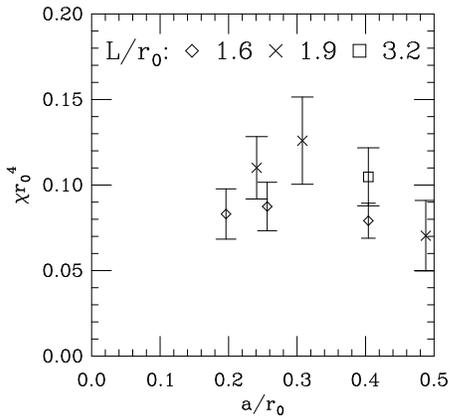}
\vskip -10mm
\end{center}
\caption{
Scaling test for topological charge,
as a function of lattice spacing. The three lattice sizes (in units of
$r_0$)
of 1.6, 1.9, and 3.2 are shown by diamonds, crosses, and a square,
respectively.
}
\label{fig:chir04}
\end{figure}

Figure \ref{fig:chir04} summarizes our results.  We plot the quantity
$\chi r_0^4$ versus the
lattice spacing $a/r_0$ for physical volumes $L/r_0 \sim 1.6,1.9$ and
$3.2$. The three data points corresponding to  $L/r_0 \approx 1.6$, 
which span a range in lattice spacing $a=0.4r_0 - 0.2r_0$ is
consistent with scaling. Apparently a lattice with $a=0.4r_0$ is fine
enough to support the physically relevant instantons.

Making the assumption that the largest volume is
large enough to approximate infinity, and making the
assumption that the pure gauge SU(2)
$r_0$ is equal to the phenomenological SU(3) number
0.5 fm, our value $\chi r_0^4= 0.12(2)$ corresponds
to  $\chi$ = (235(10) MeV)${}^4$.
Evaluating the Witten-Veneziano formula with the known physical masses
of
the appropriate mesons, and $N_f=3$,
this susceptibility yields an $\eta'$ mass of 1520 MeV.

\section{ THE TOPOLOGICAL SUSCEPTIBILITY AT $T\ne 0$} 

Not much is known about the role of instantons and the topological
susceptibility at finite temperature. At  $T<T_c$ one expects a very
weak T dependence while at very high temperatures
($T>T_c/3$) the instantons are exponentially suppressed.

\begin{figure}[htb]
\begin{center}
\vskip -20mm
\leavevmode
\epsfxsize=60mm
\epsfbox[40 50 530 590]{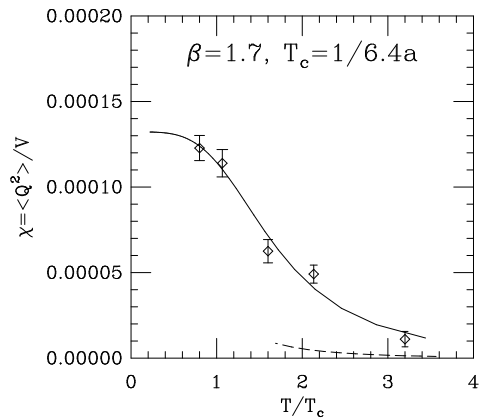}
\vskip -10mm
\end{center}
\caption{
Topological susceptibility at finite temperature. The dashed line line
is the high temperature
prediction of Yaffe and Pisarski, the solid line is a simple finite
volume model prediction.
}
\label{fig:chir_t}
\end{figure}

In our preliminary study we performed a series of simulations at a
fixed $\beta=1.7$ value corresponding to $N_T\sim 6.4$. We kept the
spatial volume fixed at $N_s=8$ and varied the temporal direction between
$N_T=8$ and $N_T=2$. Figure  \ref{fig:chir_t} shows the susceptibility
as a function of $T/T_c$. The dashed line is the high temperature
prediction of Yaffe and Pisarski while the solid line is a very simple
model prediction following Shuryak's picture assuming that instantons
with diameter approximately equal to the inverse temperature are
suppressed\cite{INSTANTON_T}.  The data points agree surprisingly well
with this simple two-parameter model prediction.
 
This work was supported by the U.S. Department of 
Energy and by the National Science Foundation.

\end{document}